\documentclass[USenglish,twocolumn]{article}

\usepackage[utf8]{inputenc}				
\usepackage[T1]{fontenc}
\usepackage[big,online]{dgruyter}	
\hypersetup{colorlinks,allcolors=blue}

\usepackage{lmodern}
\usepackage{microtype}
\usepackage[table,usenames,dvipsnames]{xcolor}
\usepackage[numbers,square,sort&compress]{natbib}

\usepackage{bm} 
\usepackage{mathbbol} 

\usepackage{comment}

\renewcommand{\Im}{\operatorname{Im}}

\newcommand{\dd}{\mathrm{d}}

\newcommand{\JJ}{\mathcal{J}}

\newcommand{\bJJ}{\boldsymbol{\JJ}}
\newcommand{\bJJmod}{\bJJ_{\mathrm{mod}}}
\newcommand{\Heff}{\tilde{\mathbf{H}}}

\newcommand{\rr}{\mathbf{r}}
\newcommand{\rrp}{\rr'}
\newcommand{\nn}{\mathbf{n}}

\newcommand{\ff}{\hat{\mathbf{f}}_{\lambda}(\rr,\omega)}
\newcommand{\ffdag}{\hat{\mathbf{f}}_{\lambda}^{\dagger}(\rr,\omega)}
\newcommand{\ffp}{\hat{\mathbf{f}}_{\lambda}(\rrp,\omega)}

\newcommand{\Gap}{\text{Gap}}
\newcommand{\Top}{\text{Top}}
\newcommand{\Bottom}{\text{Bottom}}

\begin{document}

\def\equationautorefname~#1\null{Eq.~(#1)\null}
\renewcommand{\figureautorefname}{Fig.}

	\articletype{Research Article}

\title{Few-mode Field Quantization for Multiple Emitters}
\runningtitle{Few-mode Field Quantization for Multiple Emitters}

\author[1]{Mónica Sánchez-Barquilla}
\author[1,2]{Francisco J. García-Vidal}
\author[1]{Antonio I. Fernández-Domínguez}
\author[1]{Johannes Feist*}
\runningauthor{M.~Sánchez-Barquilla et al.}
\affil[1]{\protect\raggedright{} 
Departamento de Física Teórica de la Materia Condensada and
Condensed Matter Physics Center (IFIMAC), Universidad Autónoma de Madrid,
E-28049 Madrid, Spain, e-mail: johannes.feist@uam.es}
\affil[2]{\protect\raggedright{} 
Institute of High Performance Computing, Agency for Science, Technology,
and Research (A*STAR), Connexis, 138632 Singapore}

	
\abstract{The control of the interaction between several quantum emitters using
nanophotonic structures holds great promise for quantum technology applications.
However, the theoretical description of such processes for complex
nanostructures is a highly demanding task as the electromagnetic (EM) modes are
in principle described by a high-dimensional continuum. We here introduce an
approach that permits a quantized description of the full EM field through a
``minimal'' number of discrete modes. This extends the previous work in
Ref.~\cite{Medina2021} to the case of an arbitrary number of emitters with
arbitrary orientations, without any restrictions on the emitter level structure
or dipole operators. We illustrate the power of our approach for a model system
formed by three emitters placed in different positions within a
metallodielectric photonic structure consisting of a metallic dimer embedded in
a dielectric nanosphere. The low computational demand of this method makes it
suitable for studying dynamics for a wide range of parameters. We show that
excitation transfer between the emitters is highly sensitive to the properties
of the hybrid photonic-plasmonic modes, demonstrating the potential of such
structures for achieving control over emitter interactions.}


\maketitle
	
\section{Introduction} 

The control of photon-mediated interactions between quantum emitters has
generated great interest over the last years, since it is essential for quantum
technology applications such as quantum networking, quantum information and
quantum computation~\cite{Reiserer2015, Nguyen2019, Gonzalez-Tudela2015,
Douglas2015}. Nanophotonic devices with subwavelength light confinement are
promising platforms to engineer such interactions, as the large confinement
enables large emitter-photon coupling strengths and thus fast dynamics. At the
same time, achieving strongly subwavelength confinement typically relies on the
use of highly lossy constituents such as metallic nanoparticles with plasmonic
resonances~\cite{Khurgin2015}, and furthermore requires that the quantum
emitters are brought close to the material surfaces. In these conditions, the EM
mode spectrum typically contains a series of broad and overlapping
resonances~\cite{Li2016Transformation}.

Quantizing the electromagnetic (EM) field in such systems is highly nontrivial,
as losses cannot be neglected nor treated perturbatively, such that standard
approaches of quantization fail~\cite{Glauber1991,Cohen-Tannoudji1997}. One
powerful framework that overcomes these limitations is given by macroscopic
quantum electrodynamics (QED)~\cite{Fano1956, Huttner1992, Scheel1998,
Knoll2001, Scheel2008, Buhmann2012I, Buhmann2012II, Feist2020}. It provides a
recipe for quantizing the medium-assisted EM field in arbitrary material
structures whose response is approximated through the macroscopic Maxwell's
equations, including dispersive and lossy materials. However, within this
quantization scheme, the quantized EM field is described by an extremely large
continuum of bosonic modes~\cite{Feist2020}. While this approach has proven
hugely successful for treating problems where the EM modes are treated
perturbatively or integrated out in some other way, it is not directly useful
for applying cavity-QED-like approaches in which the modes are treated as
explicit degrees of freedom of the system.

This shortcoming combined with the increasing interest in metallic and
metallodielectric subwavelength cavity QED systems, in which highly lossy
resonances act as effective cavity modes, has led to the development of several
approaches that allow the construction of few-mode quantized models in which the
full quantized EM field is approximately described through a (small) collection
of discrete quantized modes. One approach relies on quasinormal mode theory for
Maxwell's equations~\cite{Sauvan2013,Kristensen2014,Lalanne2018}, based on
forming superpositions of the modes of macroscopic QED that correspond to the
quasinormal modes (resonances) of the material structure, and then performing
appropriate approximations to obtain the energies, decay rates, and coherent and
incoherent interactions of these modes~\cite{Franke2019, Ren2021, Franke2021}.
This approach is powerful, but relies on being able to select just a few
quasinormal modes of the system. In the case of nanometric-sized metallic
structures, a similar quantization strategy has been used, but greatly
simplified through the quasi-static description of sub-wavelength-confined
plasmonic fields~\cite{Delga2014, Li2016Transformation,
Cuartero-Gonzalez2020Dipolar, Fregoni2021Strong}. An alternative approach is
obtained by exploiting the fact that, due to its linearity, a system of harmonic
oscillators (such as EM modes) is fully determined by its linear response to the
quantum emitter, encoded in the so-called spectral density. This offers the
possibility to construct a model system that shows the same response, but is
(significantly) easier to solve than the original problem. It has recently been
shown that a discrete collection of interacting modes coupled to Markovian
background baths provides exactly such a model with sufficient flexibility to
reproduce even complex responses faithfully~\cite{Medina2021}. One important
advantage of this model lies in the fact that it can account for a peak
consisting of many underlying resonances (quasinormal modes) with just one or a
few model modes. Such a situation is often encountered in the so-called
pseudomodes in plasmonic systems, which arise due to the collective response of
high-k modes in planar systems~\cite{Gonzalez-Tudela2014} or high-order
multipoles in spherical ones~\cite{Delga2014}. 

While the model developed in Ref.~\cite{Medina2021} can treat arbitrary photonic
structures, in the formulation presented therein, it is only suitable for
situations where only a single emitter is present in the system and all
considered emitter dipole transitions are co-aligned. In the present article, we
lift these restrictions and extend the approach to a collection of emitters with
arbitrary orientations of the transition dipole moments. We achieve this by
first generalizing the definition of the spectral density to the case of several
light-matter interaction operators~\cite{Cuartero-Gonzalez2018,
Yang2019Signatures, Cuartero-Gonzalez2021Distortion}. The spectral density
$J(\omega)$, which is normally a scalar function that fully characterizes the
interaction between a quantum system and a bath mediated by a single interaction
operator~\cite{Breuer2007, Mascherpa2020}, then becomes an $M\times M$
matrix-valued function. Here, $M$ is the number of distinct interaction
operators that are treated ($M = 3M_e$ for $M_e$ dipolar emitters with all three
possible dipole orientations taken into account for each emitter). We then
extend the few-mode quantization approach presented in Ref.~\cite{Medina2021} to
this case. We show that the few-mode quantization of arbitrary generalized
spectral densities can be obtained for several emitters placed at different
positions through a simple fitting procedure.

We then apply the approach to study energy transfer between emitters for three
different situations: (i) transfer of a single excitation from a coherent
superposition of two emitters to a third one; (ii) transfer of a single
excitation from one emitter to another, mediated by the third one; (iii)
excitation transfer to a third emitter from a doubly excited state where the
other two emitters are initially excited. Our method is able to calculate the
dynamics for these different examples at low computational cost. They show that
the use of metallodielectric structures allows great control in the population
transfer between emitters close to resonance to a hybrid mode, with slight
changes in the emitter parameters inducing qualitatively different dynamics. 

\section{Theory} 

We start by discussing a general model consisting of a matter part (which can
represent multiple emitters) linearly coupled to a collection of bosonic modes
(which will later represent the medium-assisted EM field). We set $\hbar=1$ here
and in the following, and write the Hamiltonian as
\begin{align}
    H = H_\mathrm{mat} + \vec A^{\dagger T} \mathbf{H} \vec A + \left(\vec{V}^T \cdot \mathbf{M} \cdot \vec A + \mathrm{H.c.}\right), \label{eq:HFano}
\end{align}
where $H_\mathrm{mat}$ describes the matter (the emitters), $\vec A = (a_1, a_2,
\ldots, a_\alpha, \ldots)^T$ collects all EM modes, and $\vec{V} = (V_1, V_2,
\ldots, V_n, \ldots)^T$ collects the emitter operators describing the
interaction with the bosonic modes. In the case considered below, $\vec{V}$ will
contain the dipole operators (up to three per emitter if all polarizations have
to be taken into account). The properties of the bosonic environment are thus
fully encoded in the matrices $\mathbf{H}$ (size $N_a\times N_a$) and
$\mathbf{M}$ (size $M\times N_a$), where $N_a$ is the number of bosonic modes
and $M$ is the number of matter interaction operators. We note that for
simplicity of notation we discuss a formally discretized bosonic bath and will
perform the continuum limit $N_a\to\infty$ when required.

We now define the generalized spectral density associated to the bosonic
environment in \autoref{eq:HFano} as
\begin{equation}\label{eq:genJ}
  \bJJ(\omega) = \lim_{\epsilon\to0} \frac{1}{\pi} \mathbf{M} \Im\left[\frac{1}{\mathbf{H} - \omega - i\epsilon}\right] \mathbf{M}^\dagger.
\end{equation}
This definition is a straightforward extension of the single-emitter spectral
density to the case of multiple light-matter interaction operators, obtained by
replacing a $1\times N_a$ vector of light-matter coupling elements by the
$M\times N_a$ matrix $\mathbf{M}$, and has been previously obtained in the
context of the Wigner-Weisskopf problem (i.e., within the single-excitation
subspace)~\cite{Cuartero-Gonzalez2018, Yang2019Signatures,
Cuartero-Gonzalez2021Distortion}. The generalized spectral density is an
$M\times M$ matrix-valued function of frequency and fulfills
$\bJJ(\omega)^\dagger = \bJJ(\omega)$. We note that when the emitters are
approximated as two-level systems, the transition dipole moments are usually
included in $\mathbf{M}$, such that the interaction operators $V_n$ become
unitless and $\bJJ(\omega)$ has units of frequency~\cite{Cuartero-Gonzalez2018}.

If $\mathbf{H}$ is diagonal, $H_{\alpha\beta} = \omega_\alpha
\delta_{\alpha\beta}$, we can use the Sokhotski-Plemelj formula
$\lim_{\epsilon\to0} \frac{1}{\omega' - \omega - i\epsilon} = \mathcal{P}
\frac{1}{\omega' - \omega} + i\pi \delta(\omega'-\omega)$ to get
\begin{equation}\label{eq:genJ_elem}
  \JJ_{nm}(\omega) = \sum_{\alpha} M_{n\alpha} M^*_{m\alpha} \delta(\omega - \omega_\alpha),
\end{equation}
which is a form where the relation to conventional single-emitter spectral
densities $J(\omega) = \sum_\alpha |M_\alpha|^2 \delta(\omega-\omega_\alpha)$
appears even more clearly. We show below that $\bJJ(\omega)$ indeed encodes the
full information about the environment that the emitters interact with.

To connect the general Hamiltonian \autoref{eq:HFano} to the physical system we
are interested in (a collection of emitters interacting with the EM field
supported by a material structure), we use the framework of macroscopic QED\@.
The Hamiltonian in the multipolar coupling scheme (Power-Zienau-Woolley picture)
and within the dipole approximation can then be written as
\begin{multline}\label{eq:Hmqed}
    H = \sum_{\lambda}\int \dd^3r \int_0^{\infty} \dd\omega\, \omega \ffdag \ff \\
       + \sum_k H_k - \sum_k \bm{\mu}_k \cdot \hat{\mathbf{E}}(\rr_k),
\end{multline}
where $\lambda = \{e,m\}$ labels the electric and magnetic contributions, $M$ is
the number of emitters, $\ff$ and $\ffdag$ are the bosonic annihilation and
creation operators of the medium-assisted field, and $H_k$ and $\bm{\mu}_k$ are
the bare Hamiltonian and dipole operator of emitter $k$. The electric field
operator is given by
\begin{equation}\label{eq:Emqed}
  \hat{\mathbf{E}}(\rr) = \sum_{\lambda}\int \dd^3r' \int_0^{\infty} \dd\omega \mathbf{G}_{\lambda}(\rr,\rrp,\omega)\cdot \ffp + \text{H.c.},
\end{equation}
where $\mathbf{G}_{\lambda}(\rr,\rrp,\omega)$ are the electric and magnetic
Green's functions, which are closely related to the dyadic Green's
function~\cite{Scheel2008}. This Hamiltonian can be rewritten in the form of
\autoref{eq:HFano} by formally discretizing space and frequency and defining
$a_\alpha = \nn_\alpha \cdot
\hat{\mathbf{f}}_{\lambda_\alpha}(\rr_\alpha,\omega_\alpha)$, with a combined
mode index $\alpha \equiv (\lambda_\alpha, \rr_\alpha, \omega_\alpha,
\nn_\alpha)$, where $\nn_\alpha \in \{\hat{\mathbf{x}}, \hat{\mathbf{y}},
\hat{\mathbf{z}}\}$. Furthermore, the interaction operators $V_n = \nn_n\cdot
\bm{\mu}_{k_n}$ are determined by a combined index $n \equiv (k_n,\nn_n)$, where
the unit vectors $\nn_n$ give the (up to three) dipole directions taken into
account for each emitter. This way, we can identify $H_{\alpha\beta} =
\delta_{\alpha\beta} \omega_\alpha$ and $M_{n\alpha} = -\nn_n \cdot
\mathbf{G}_{\lambda_\alpha}(\rr_n,\rr_\alpha,\omega_\alpha) \cdot \nn_\alpha$.

Inserting the expression for $M_{n\alpha}$ in \autoref{eq:genJ_elem}, taking the
continuum limit (i.e., replacing the sum over $\alpha$ with the corresponding
sums and integrals), and using the Green's function integral
identity~\cite{Scheel2008}
\begin{equation}
  \sum_{\lambda}\!\int\!\dd^3\mathbf{s}\mathbf{G}_{\lambda}(\rr,\mathbf{s},\omega)\cdot \mathbf{G}_{\lambda}^{*T}(\rrp,\mathbf{s},\omega) = \frac{\omega^2}{\pi\epsilon_0 c^2} \Im \mathbf{G}(\rr,\rrp,\omega)
\end{equation}
leads to
\begin{equation}\label{eq:genJ_MQED}
    \JJ_{nm}(\omega) = \frac{\omega^2}{\pi\epsilon_0 c^2} \nn_n \cdot \Im \mathbf{G}(\rr_n,\rr_m,\omega) \cdot \nn_m.
\end{equation}
The diagonal elements of $\bJJ(\omega)$ are equal to the ``conventional''
spectral density for a single transition up to the square of the dipole
transition matrix element, $\mu_n^2 \JJ_{nn}(\omega) = J_n(\omega)$. From
\autoref{eq:genJ_MQED}, it can be seen that $\bJJ(\omega)$ is a real and
symmetric matrix for any frequency $\omega$, and due to the properties of the EM
dyadic Green's function, it is also positive definite. This means that it can be
decomposed in the form $\bJJ(\omega) = \mathbf{g}(\omega)
\mathbf{g}(\omega)^{\dagger}$, where $\mathbf{g}(\omega)$ can be chosen real and
is unique up to unitary transformations $\mathbf{g}'(\omega) =
\mathbf{g}(\omega) \mathbf{U}(\omega)$. We now note that $\mathbf{g}(\omega)$ is
exactly the coupling matrix that appears in the expansion of the multi-emitter
problem using emitter-centered modes in macroscopic QED~\cite{Feist2020}. As
discussed and demonstrated in that reference, this quantity indeed encodes the
full information about the EM environment, and consequently, so does
$\bJJ(\omega)$. This means that two different systems with the same
$\bJJ(\omega)$ are indistinguishable from the point of view of the emitters. We
note in passing that $\bJJ(\omega)$ is in fact a more fundamental quantity than
$\mathbf{g}(\omega)$, as it does not depend on an arbitrary choice of basis.

We now extend the model presented in~\cite{Medina2021} in order to obtain an
effective few-mode description of the multi-emitter problem. As discussed there,
the idea is to find a model system that is equivalent to the actual EM
environment, but has a structure that facilitates its numerical solution and
interpretation of the resulting dynamics. We again introduce a discrete set of
$N$ mutually coupled discrete EM modes $a_i$, each of which is coupled to an
independent bath of ``background'' modes $b_{i,\Omega}$ with
frequency-independent coupling determined by $\nu_i = \sqrt{\kappa_i/(2\pi)}$,
and also to the emitter interaction operator $V_n$ with coupling strength
$g_{ni}$. The background bath modes are not directly coupled to the emitters.
The model Hamiltonian is then given by $\mathcal{H} = \mathcal{H}_S +
\mathcal{H}_B$, where
\begin{subequations}
\begin{align}
  \mathcal{H}_S &= \sum_n H_n + \sum_{i,j} \omega_{ij} a^\dagger_{i} a_{j} + \sum_{n,i} V_n g_{ni} (a_i + a_i^\dagger),\label{eq:H_S}\\
  \mathcal{H}_B &= \sum_{i} \int\left[\Omega b_{i,\Omega}^\dagger b_{i,\Omega} + \nu_i (b_{i,\Omega}^\dagger a_i + b_{i,\Omega} a_i^\dagger)\right]
  \mathrm{d}\Omega,
\end{align}
\end{subequations}
Since the coupling of the discrete modes to the background baths is spectrally
flat, it is Markovian. The dynamics of the system are then
equivalently described~\cite{Mascherpa2020} by the Lindblad master equation
\begin{equation}
  \dot{\rho} = -i\left[\mathcal{H}_S, \rho\right] + \sum_i \kappa_i L_{a_i}[\rho],
\end{equation}
where $\rho$ is the system density matrix and $L_O[\rho] = O \rho O^{\dagger}
-\frac12\{O^{\dagger}O,\rho\}$ is a Lindblad dissipator.

The model Hamiltonian $\mathcal{H}$ can also be rewritten in the form of
\autoref{eq:HFano}. To do so, we formally discretize the bath continua, with
$N_b$ modes for each continuum, such that there are $N_a = N(N_b+1)$ bosonic
modes, given by $\vec{A} = \left(a_1, \dots, a_N, b_{1,\Omega_1}, \dots,
b_{1,\Omega_{N_b}}, b_{2,\Omega_1}, \dots, b_{N,\Omega_{N_b}}\right)^T$. In this
form, $\mathbf{H}$ is not diagonal, but consists of the block matrix
$\omega_{ij}$ in the top left, a series of diagonal blocks for each continuum,
and constant off-diagonal coupling elements between all modes of a block and the
discrete mode associated to it. Finally, $\mathbf{M}$ is an $M\times N(N_b+1)$
matrix in block form, $\mathbf{M} = \begin{pmatrix}\mathbf{g} &
\mathbf{0}\end{pmatrix}$, where $\mathbf{g}$ is the real $M\times N$ matrix
containing the coupling elements $g_{ni}$.

\begin{figure*}[tb]
  \includegraphics[width=\linewidth]{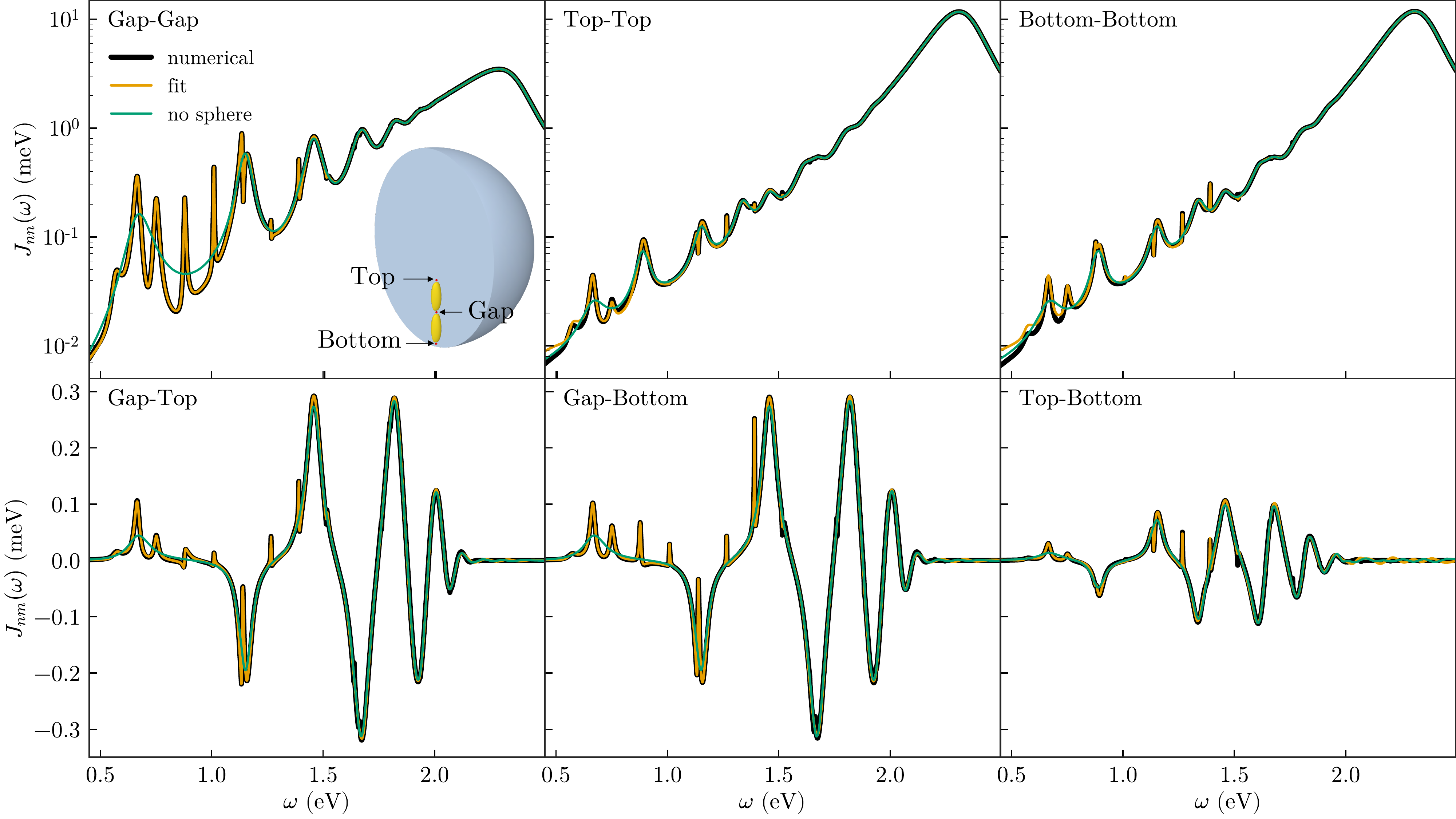}
  \caption{Generalized spectral densities for $z$-oriented emitters at Gap, Top
  and Bottom positions (thick black line), fitted model spectral density (orange
  line), and spectral density when the microsphere is replaced by a dielectric
  background (green line). Inset: Sketch of the system consisting of a silver
  dimer nanoantenna embedded in a dielectric microsphere (with the same
  dimensions as in~\cite{Medina2021}). The red dots show the position of each
  emitter.}
  \label{fig:Jijs}
\end{figure*}

A compact form of the generalized spectral density of the model system can be
obtained from \autoref{eq:genJ}, either by explicit diagonalization of
$\mathbf{H}$ using the Lippmann-Schwinger formalism as in the supplemental
material of Ref.~\cite{Medina2021}, or by following the approach of
Ref.~\cite{Glutsch2002}. The resulting expression is
\begin{equation}\label{eq:JJmodel}
  \bJJmod(\omega) = \frac{1}{\pi} \mathbf{g} \Im\left[\frac{1}{\Heff - \omega} \right] \mathbf{g}^T,
\end{equation}
where $\Heff$ is a complex symmetric $N\times N$ matrix with elements given by
$\Heff_{ij} = \omega_{ij}-\frac{i}{2} \kappa_i\delta_{ij}$. When $\Heff$ is
diagonalizable, $\Heff = \mathbf{V} \tilde{\mathbf{\Omega}} \mathbf{V}^T$, where
$\tilde{\mathbf{\Omega}}$ is a diagonal matrix containing the (complex)
eigenvalues and $\mathbf{V}$ is a complex orthogonal (not unitary) matrix,
$\mathbf{V}^T \mathbf{V} = \mathbf{1}$, this expression can be rewritten as a
sum over resonances,
\begin{equation}\label{eq:JJmodel_diag}
  \bJJmod(\omega) = \frac{1}{\pi} \Im\left[\tilde{\mathbf{g}} \frac{1}{\tilde{\mathbf{\Omega}} - \omega} \tilde{\mathbf{g}}^T\right],
\end{equation}
where $\tilde{\mathbf{g}} = \mathbf{g}\mathbf{V}$. The complex eigenvalues of
$\Heff$ are thus the poles of the spectral density, and closely related to the
quasinormal modes of the classical Green's function. This form also facilitates
the classification of multi-mode effects~\cite{Lentrodt2021}.

As in the single-emitter case, the relatively few parameters in
\autoref{eq:JJmodel} can be adjusted to best reproduce the generalized spectral
density \autoref{eq:genJ_MQED} calculated from the dyadic EM Green's function,
and thus parametrize the model Hamiltonian $\mathcal{H}$. The symmetry of
$\bJJ(\omega)$ means that this corresponds to a simultaneous fit of $M(M+1)/2$
real-valued functions. The off-diagonal elements of the generalized spectral
density, which encode the interaction between the emitters mediated by the EM
fields, can have negative values, while the diagonal elements, which correspond
to the ``conventional'' spectral densities, are non-negative, i.e.,
$J_{nn}(\omega)\ge 0$ for all $\omega$. We note that the spectral density is
fully determined by the position of the emitters, with no restriction on the
emitter properties. In particular, this approach is not restricted to two-level
systems. However, if only one or two dipole orientations are relevant for the
transitions of any emitter, the number of necessary interaction operators and
thus the dimensionality of $\bJJ(\omega)$ can be reduced.

For completeness, we note that the number of free parameters in the diagonal
form given in \autoref{eq:JJmodel_diag} is typically significantly smaller than
in the non-diagonal form given in \autoref{eq:JJmodel}. In the non-diagonal
form, the number of real parameters needed is $\frac12 N(N+1)$ for
$\omega_{ij}$, $N$ for $\kappa_i$, and $N M$ for $g_{ni}$, giving a total of
$\frac12 N(N+2M+3)$ free real parameters. In the diagonal form, there are $N$
complex parameters $\tilde{\Omega}_i$, while $\tilde{\mathbf{g}}$ has $N M$
complex parameters that are restricted by up to $\frac12 M(M+1)$ relations since
$\tilde{\mathbf{g}} \tilde{\mathbf{g}}^T = \mathbf{g} \mathbf{g}^T$ must be
purely real, giving $\frac12 (4N-M)(M+1)$ real parameters (when $N>M$). It would
thus seem that this form is more convenient for fitting than the non-diagonal
form. However, this turns out not to be the case~\cite{Sanchez-Barquilla2021}:
First, it is not straightforward to enforce that the fit parameters correspond
to a physical system where $\bJJmod(\omega)$ is real positive semidefinite for
all frequencies. Second, even when that constraint is achieved, implementation
of the dynamics through a Lindblad master equation (which as discussed above is
the final goal of this approach) requires a form in which the imaginary part of
the complex symmetric matrix $\Heff$ is negative semidefinite, while
$\mathbf{g}$ is real. This would require an algorithm that finds a complex
orthogonal matrix $\mathbf{V}$ which ``undiagonalizes'' the system and can be
used to obtain physical $\Heff = \mathbf{V} \tilde{\mathbf{\Omega}}
\mathbf{V}^T$ and $\mathbf{g} = \tilde{\mathbf{g}} \mathbf{V}^T$ for any given
$\tilde{\mathbf{\Omega}}$ and $\tilde{\mathbf{g}}$. To the best of our
knowledge, no algorithm that achieves this has been found~\cite{Pleasance2020,
Sanchez-Barquilla2021}. It is thus more straightforward to directly use the
non-diagonal form \autoref{eq:JJmodel} when fitting.

\section{Results}

To illustrate the generalization to the multiple emitter case, we consider the
same physical setup as in Ref.~\cite{Medina2021}, see the inset in
\autoref{fig:Jijs}. It consists of two silver nanoparticles (ellipsoids with
long axis of $120$~nm and short axis of $40$~nm), separated by a $3$~nm gap, and
embedded in a dielectric GaP ($\epsilon_{\text{sph}}=9$) nanosphere of $600$~nm
radius, with the rods substantially displaced from the center of the sphere. We
consider three two-level emitters placed in different positions, indicated by
red dots in the inset in \autoref{fig:Jijs}. We will refer to the emitter
positions as: (i) Gap (in the center between the rods), (ii) Top ($1.5$~nm above
the upper rod) and (iii) Bottom ($1.5$~nm below the lower rod). The generalized
spectral density then is characterized by six independent functions (three
diagonal and three off-diagonal). Since we consider two-level quantum emitters
with a single dipole transition, we include the transition dipole moments within
the spectral density for simplicity, $J_{nm}(\omega) = \mu_n \mu_m
\JJ_{nm}(\omega)$, with values given by $\mu_\Top = \mu_\Bottom = 0.55$~e\,nm
and $\mu_\Gap = 0.257$~e\,nm. The interaction operators are then just $V_n =
\sigma_n^+ + \sigma_n^-$. \autoref{fig:Jijs} shows the generalized spectral
density (black lines) obtained from classical simulations performed with the
Maxwell equations solver implemented in COMSOL Multiphysics. The first row shows
the diagonal functions (in logarithmic scale), i.e., $J_{nn}(\omega)$, which
correspond to the ``conventional'' spectral densities, while the second row
shows the off-diagonal functions, $J_{12}(\omega)$, $J_{13}(\omega)$, and
$J_{23}(\omega)$, which encode the field-mediated interaction between the
emitters. As Top and Bottom are symmetric positions with respect to the
plasmonic nanoparticles, both their diagonal elements and their interaction with
the emitter in Gap behave quite similarly, with slight differences due to the
nonsymmetric placement of the dielectric sphere.

In its most general form, there are no restrictions on the form of $\Heff$ apart
from symmetry. However, it is possible to choose any desired structure, e.g.,
inspired by the physical structure of the problem, to restrict the number of
free parameters. In Ref.~\cite{Sanchez-Barquilla2021}, we recently showed that
for the one-emitter case, it is generally sufficient to use a chain form with
only next-nearest neighbor coupling between the modes (i.e., $\Heff_{ij}=0$ if
$|i-j|>2$), although this choice makes it more challenging to obtain converged
fits. We here instead choose a block-diagonal form,
\begin{equation}
    \Heff = \begin{pmatrix}
        \Heff_1 & \mathbf{0} & \mathbf{0} \\
        \mathbf{0} & \Heff_2 & \mathbf{0} \\ 
        \mathbf{0} & \mathbf{0} & \Heff_3 
    \end{pmatrix},
\end{equation}
where $\Heff_i$ are full $N_i\times N_i$ matrices. This form allows to
significantly decrease the number of parameters while still giving a good fit
for the spectral density. The physical reasoning behind this election is that
there are many independent modes in the system that do not interfere
significantly (e.g., the high-order ``pseudomodes'' coupling to each
emitter~\cite{Delga2014}), and it is thus not necessary to allow arbitrary
couplings between all modes. In the present case, the size of each block was
chosen as $N_i=14$, giving a total of $N=\sum_{i} N_i = 42$ modes.

\begin{figure*}[tb]
  \includegraphics[width=\linewidth]{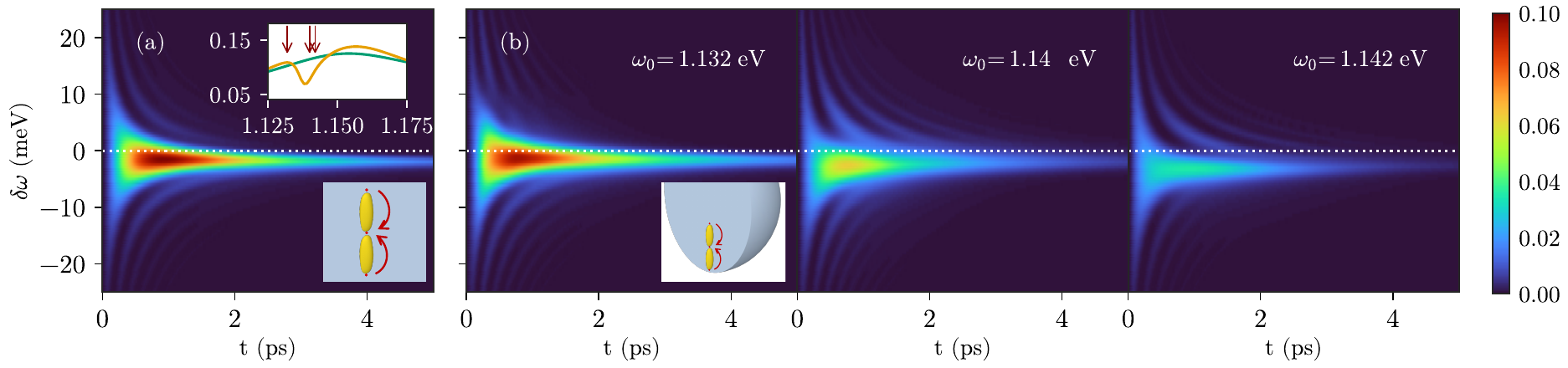}
  \caption{Population transfer to Gap when Top and Bottom are fixed at a
  frequency $\omega_0$ and Gap is detuned from that frequency. The initial state
  $\psi_0$ is a superposition of Top and Bottom excited, as shown in both
  sketches  in subplots (a) for the dielectric background and (b) for the
  dielectric sphere, where three different frequencies are considered (shown in
  each subplot). The upper inset in subplot (a) shows the spectral density of the 
  Gap emitter around the hybrid mode for the dielectric sphere (orange line) and the
  dielectric background (green line).}
  \label{fig:popGap}
\end{figure*}

The orange lines in \autoref{fig:Jijs} show the model spectral density obtained
after fitting. Despite the complexity of the structure and the spectral
densities, the fit is well-converged with a relatively small number of modes. 
The small differences to the numerical spectral density visible at low
frequencies for the diagonal functions and at high frequencies for the
non-diagonal ones could be reduced by increasing the number of modes employed in
the fit, but were found not to affect the dynamics studied in this work.

In \autoref{fig:Jijs}, we additionally show the spectral density corresponding
to the two plasmonic nanoparticles when the sphere is replaced by an infinite
GaP background medium (green lines). In this case, Top and Bottom are completely
equivalent positions, and their spectral densities are identical. Additionally,
this change removes the Mie resonances supported by the sphere, such that the
spectral density is overall much simpler and contains fewer peaks. In
particular, there are no visible interference structures, and the spectral
density corresponds to a series of broad but mostly independent modes. The
fitting procedure then converges much more easily and achieves even better
agreement with the numerical results, with $|\bJJmod(\omega)-\bJJ(\omega)| <
0.015~$meV over the full spectral range of \autoref{fig:Jijs} using $N=30$
modes. As the fit is visually indistinguishable from the exact spectral density,
it is not shown separately in \autoref{fig:Jijs}.

We now study the energy transfer dynamics between the emitters, with a focus on
how it is influenced by the formation of hybrid modes. We choose emitters with
frequencies close to the lowest-energy hybrid modes at $\approx 1.14$~eV. The
upper inset in \autoref{fig:popGap}(a) shows the spectral density of the Gap
emitter in that frequency range, both for the hybrid metallodielectric cavity
(orange line) and for the plasmonic dimer embedded in an infinite dielectric
medium (green line). For the hybrid cavity, there is significant mode
hybridization and destructive interference around that frequency, while for the
bare dimer, only a single broad resonance peak appears.

\begin{figure*}[tb]
  \includegraphics[width=\linewidth]{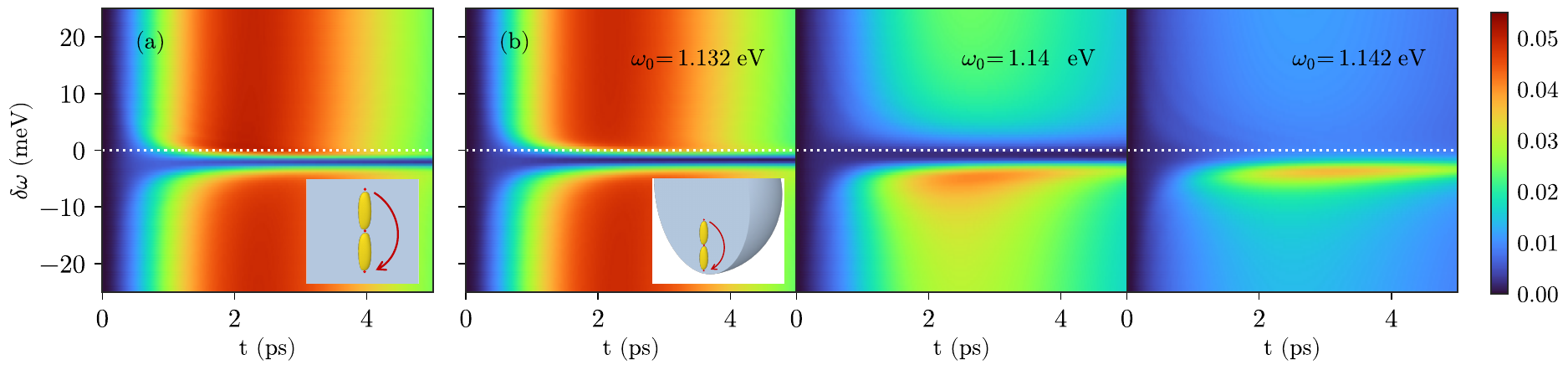}
  \caption{Population transfer to Bottom when Top and Bottom are fixed at a
  frequency $\omega_0$ and Gap is detuned from that frequency. The initial state
  $\psi_0$ corresponds to only Top excited, as shown in both sketches in
  subplots (a) for the dielectric background and (b) for the dielectric sphere
  (shown in each subplot).}
  \label{fig:popBottom}
\end{figure*}

We first investigate population transfer from a coherent superposition of the
Top and Bottom emitters to the Gap emitter, as schematically shown in the lower
insets of \autoref{fig:popGap}. To be precise, we calculate the dynamics for an
initial state $|\psi_0\rangle = \frac{1}{\sqrt{2}} (\sigma_{\Top}^+ +
\sigma_{\Bottom}^+) |0\rangle$. We fix the frequencies of the Top and Bottom
emitters to be equal, $\omega_\Top=\omega_\Bottom=\omega_0$ and vary the
frequency of the Gap emitter, $\omega_\Gap = \omega_0 + \delta\omega$. The
population $P_\Gap(t) = \langle \sigma_\Gap^+ \sigma_\Gap^-\rangle$ as a
function of time and $\delta\omega$ is shown for three distinct values of
$\omega_0$, indicated by the dark red arrows in the upper inset of
\autoref{fig:popGap}. Panel (a) corresponds to the case of a dielectric
background, for which we only consider $\omega_0 = 1.142~$eV (the results for
the other two values are very similar due to the broad nature of the peak). We
find significant excitation of the Gap emitter for a narrow range of
frequencies, which however does not coincide with $\delta\omega=0$ as could be
naively expected. This is due to the fact that the EM modes induce a significant
Lamb shift on the emitters, which is larger for the Top and Bottom emitters due
to their higher dipole moments (even though the EM mode density at the Gap
position is higher due to the interaction with both ellipsoids). Their effective
frequencies are thus lowered more than that of the Gap emitter, and resonant
energy transfer is only possible when the Gap emitter is detuned to a slightly
lower frequency, $\delta\omega\approx -3$~meV.

In panel (b), we explore the situation for the hybrid cavity, where the peak
splits due to interaction between the Mie resonances of the microsphere and the
plasmonic dimer modes. Since the spectral density here has significantly more
structure, we explore the energy transfer for three values of $\omega_0$:
$1.132$~eV, $1.14$~eV, and $1.142$~eV. As shown in \autoref{fig:popGap}(b),
these small changes in frequency have a significant effect on the efficiency of
energy transfer even when the detuning is optimized to compensate for the
difference in Lamb shifts. In particular, the maximum population reaching the
Gap emitter is decreased by a factor of more than two simply when changing
$\omega_0$ by just $10$~meV. This demonstrates both the sensitivity of energy
transfer at the nanoscale to the details of the EM environment and the large
degree of control that hybrid metallodielectric structures offer for influencing
emitter dynamics. We also note that even though the emitters are not in the
regime of strong light-matter coupling (no Rabi oscillations are visible on
resonance), the effects observed here could not be reproduced with traditional
methods based on the weak-coupling approximation~\cite{Dung2002Resonant,
Gonzalez-Tudela2011, Martin-Cano2011, Downing2019}. In that approach, the EM
environment is traced out fully, with the real and imaginary parts of the
Green's function giving coherent and incoherent interactions between emitters
(with the diagonal parts corresponding to Lamb shifts and decay rates). However,
that approach is only valid when all emitter frequencies are equal (and the
emitters are two-level systems characterized by a single frequency), and
furthermore does not capture the fact that the effective frequencies of the
emitters are affected by the EM environment (and by the emitter-emitter coupling
mediated through them). Thus, the Green's function is evaluated at the bare
(unshifted) frequency. For highly structured spectral densities as in the
present case, this can be a significant source of error. We additionally note
that the effects discussed here are not correctly represented either when only
the EM modes close to resonance with the emitters are taken into account, as the
Lamb shift is dominated by off-resonant contributions.

Next, we study energy transfer from Top to Bottom, as depicted schematically in
the insets of \autoref{fig:popBottom}. We use the same system and parameters as
in the previous setup, but now initialize the system in $|\psi_0\rangle =
\sigma_\Top^+|0\rangle$, and monitor the population $P_\Bottom(t) = \langle
\sigma_\Bottom^+ \sigma_\Bottom^-\rangle$. Panel (a) in \autoref{fig:popBottom}
again shows this for the purely plasmonic nanocavity. Population transferred to
Bottom is then essentially unaffected by the presence of the Gap emitter unless
that emitter is on resonance with the others (after taking into account the
differing Lamb shifts). On resonance, energy transfer is significantly
suppressed and the Gap emitter acts like an intermediate absorber. This simple
picture is again mostly independent of $\omega_0$, so only a single value is
shown in \autoref{fig:popBottom}(a). However, in the hybrid cavity,
\autoref{fig:popBottom}(b), the picture changes drastically. In that case, the
intermediate emitter can act to either enhance or suppress the population
transfer depending on $\omega_0$ and $\delta\omega$, with a clear asymmetry with
regards to the sign of $\delta\omega$. For $\omega_0=1.14$~eV, the maximum
population reaching the Bottom emitter has a clear Fano-like interference shape
with a maximum followed by a steep minimum as a function of $\delta\omega$. When
the frequency of the Top and Bottom emitters is further increased by just
$2$~meV, to $\omega_0=1.142$~eV, the minimum essentially disappears and the
presence of the Gap emitter on resonance leads to an enhancement of energy
transfer. We thus find that within a narrow frequency range, the hybrid modes
offer the possibility to change the role of the Gap emitter from inhibiting
energy transfer between two emitters to enhancing it.

\begin{figure}[tbp]
  \includegraphics[width=\linewidth]{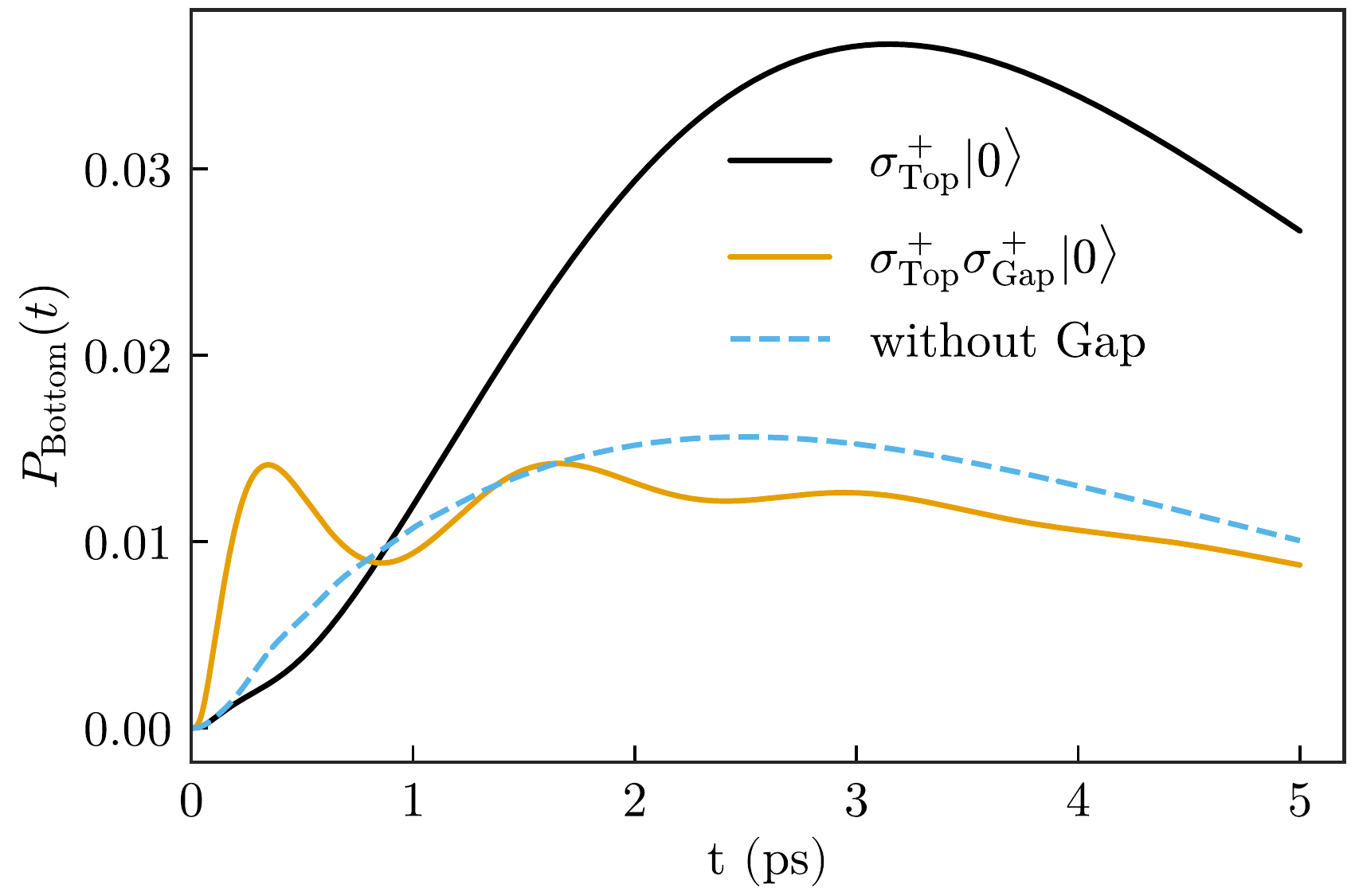}
  \caption{Population of Bottom for two-photon emission (black line) and one-photon emission (orange line) when Top and 
  Bottom have frequencies $\omega_0 = 1.142$ eV and Gap is detuned $\delta\omega=-3.78$ meV.}
  \label{fig:2photons}
\end{figure}

Finally, we study how the energy transfer from the Top to the Bottom emitter
changes when the Gap emitter is excited as well, i.e., for the initial state
$|\psi_0\rangle = \sigma_\Top^+ \sigma_\Gap^+ |0\rangle$. We here choose
$\omega_0 = 1.142$~eV, corresponding to the final dataset in
\autoref{fig:popBottom}(b), and choose the frequency of the Gap emitter so that
in the single-excitation case studied previously, the population transfer to the
Bottom emitter is maximized, achieved for $\delta\omega = -3.18$~meV.
\autoref{fig:2photons} shows $P_\Bottom(t)$ for the three cases where Gap is
initially in its ground state (black line, same data as
\autoref{fig:popBottom}), excited state (orange line), or is not present at all
(dashed blue line). When the Gap emitter is initially excited, there is fast
initial population transfer (presumably directly from the Gap to the Bottom
emitter), but the maximum population reached is significantly lower than for the
previous situation where energy transfer is enhanced by the presence of the
(ground-state) Gap emitter. In particular, when the Gap emitter is initially
excited, the maximum population in Bottom remains smaller than for the case
where no emitter is present in the Gap. This shows that the energy transfer
between the Top and Bottom emitters can be controlled by exciting the emitter in
the Gap, pointing towards a possible path for implementing quantum gates based
on hybrid metallodielectric structures.

\section{Conclusions}

In this article, we have introduced an extension of the few-mode field
quantization approach we recently developed~\cite{Medina2021} to the case of
several emitters. We have first demonstrated how to define and obtain the
generalized spectral density $\bJJ(\omega)$, a matrix-valued function that fully
determines the properties of the EM environment interacting with the emitters.
We have then shown how to obtain a few-mode quantization of the EM field in this
situation. The resulting model can be adjusted to represent arbitrary material
structures by fitting the model parameters to reproduce the numerically
calculated generalized spectral densities (which requires only the calculation
of the dyadic Green's function of Maxwell's equations and can be done with any
standard EM solver). 

We illustrated the approach in a metallodielectric structure consisting of a
metallic dimer embedded in a dielectric sphere, which produces a complex
generalized spectral density, with $N=42$ modes required to obtain a
well-converged representation of the generalized spectral density. Once the fit
is obtained, the emitter dynamics can be calculated using standard approaches of
quantum optics such as solving the Lindblad master equation. We used this to
study energy transfer between three emitters in three different situations, and
found that hybrid metallodielectric structures can enable significant control,
strongly enhancing or suppressing energy transfer for slight variations of the
emitter parameters.

\begin{acknowledgement}
  We thank Diego Martín Cano for interesting discussions about quasinormal mode
  quantization.
\end{acknowledgement}

\begin{funding}
  This work has been funded by the European Research Council
  (DOI:\doi{10.13039/501100000781}) through Grant ERC-2016-StG-714870 and by the
  Spanish Ministry for Science, Innovation, and Universities -- Agencia Estatal
  de Investigación (DOI:\doi{10.13039/501100011033}) through grants
  RTI2018-099737-B-I00, PCI2018-093145 (through the QuantERA program of the
  European Commission), and CEX2018-000805-M (through the María de Maeztu
  program for Units of Excellence in R\&D). We also acknowledge financial
  support from the Proyecto Sinérgico CAM 2020 Y2020/TCS-6545 (NanoQuCo-CM) of
  the Community of Madrid (DOI:\doi{10.13039/100012818}).
\end{funding}

\bibliographystyle{apsrev4-1}
\nocite{apsrev41Control}
\bibliography{apsrevbibopts,references}

\begin{thebibliography}{40}%
\makeatletter
\providecommand \@ifxundefined [1]{%
 \@ifx{#1\undefined}
}%
\providecommand \@ifnum [1]{%
 \ifnum #1\expandafter \@firstoftwo
 \else \expandafter \@secondoftwo
 \fi
}%
\providecommand \@ifx [1]{%
 \ifx #1\expandafter \@firstoftwo
 \else \expandafter \@secondoftwo
 \fi
}%
\providecommand \natexlab [1]{#1}%
\providecommand \enquote  [1]{``#1''}%
\providecommand \bibnamefont  [1]{#1}%
\providecommand \bibfnamefont [1]{#1}%
\providecommand \citenamefont [1]{#1}%
\providecommand \href@noop [0]{\@secondoftwo}%
\providecommand \href [0]{\begingroup \@sanitize@url \@href}%
\providecommand \@href[1]{\@@startlink{#1}\@@href}%
\providecommand \@@href[1]{\endgroup#1\@@endlink}%
\providecommand \@sanitize@url [0]{\catcode `\\12\catcode `\$12\catcode
  `\&12\catcode `\#12\catcode `\^12\catcode `\_12\catcode `\%12\relax}%
\providecommand \@@startlink[1]{}%
\providecommand \@@endlink[0]{}%
\providecommand \url  [0]{\begingroup\@sanitize@url \@url }%
\providecommand \@url [1]{\endgroup\@href {#1}{\urlprefix }}%
\providecommand \urlprefix  [0]{URL }%
\providecommand \Eprint [0]{\href }%
\providecommand \doibase [0]{http://dx.doi.org/}%
\providecommand \selectlanguage [0]{\@gobble}%
\providecommand \bibinfo  [0]{\@secondoftwo}%
\providecommand \bibfield  [0]{\@secondoftwo}%
\providecommand \translation [1]{[#1]}%
\providecommand \BibitemOpen [0]{}%
\providecommand \bibitemStop [0]{}%
\providecommand \bibitemNoStop [0]{.\EOS\space}%
\providecommand \EOS [0]{\spacefactor3000\relax}%
\providecommand \BibitemShut  [1]{\csname bibitem#1\endcsname}%
\let\auto@bib@innerbib\@empty
\bibitem [{\citenamefont {Medina}\ \emph {et~al.}(2021)\citenamefont {Medina},
  \citenamefont {{Garc{\'i}a-Vidal}}, \citenamefont
  {{Fern{\'a}ndez-Dom{\'i}nguez}},\ and\ \citenamefont {Feist}}]{Medina2021}%
  \BibitemOpen
  \bibfield  {author} {\bibinfo {author} {\bibfnamefont {I.}~\bibnamefont
  {Medina}}, \bibinfo {author} {\bibfnamefont {F.~J.}\ \bibnamefont
  {{Garc{\'i}a-Vidal}}}, \bibinfo {author} {\bibfnamefont {A.~I.}\ \bibnamefont
  {{Fern{\'a}ndez-Dom{\'i}nguez}}}, \ and\ \bibinfo {author} {\bibfnamefont
  {J.}~\bibnamefont {Feist}},\ }\bibfield  {title} {\enquote {\bibinfo {title}
  {{Few-{{Mode Field Quantization}} of {{Arbitrary Electromagnetic Spectral
  Densities}}}},}\ }\href {\doibase 10.1103/PhysRevLett.126.093601} {\bibfield
  {journal} {\bibinfo  {journal} {Phys. Rev. Lett.}\ }\textbf {\bibinfo
  {volume} {126}},\ \bibinfo {pages} {093601} (\bibinfo {year}
  {2021})}\BibitemShut {NoStop}%
\bibitem [{\citenamefont {Reiserer}\ and\ \citenamefont
  {Rempe}(2015)}]{Reiserer2015}%
  \BibitemOpen
  \bibfield  {author} {\bibinfo {author} {\bibfnamefont {A.}~\bibnamefont
  {Reiserer}}\ and\ \bibinfo {author} {\bibfnamefont {G.}~\bibnamefont
  {Rempe}},\ }\bibfield  {title} {\enquote {\bibinfo {title} {{Cavity-Based
  Quantum Networks with Single Atoms and Optical Photons}},}\ }\href {\doibase
  10.1103/RevModPhys.87.1379} {\bibfield  {journal} {\bibinfo  {journal} {Rev.
  Mod. Phys.}\ }\textbf {\bibinfo {volume} {87}},\ \bibinfo {pages} {1379}
  (\bibinfo {year} {2015})}\BibitemShut {NoStop}%
\bibitem [{\citenamefont {Nguyen}\ \emph {et~al.}(2019)\citenamefont {Nguyen},
  \citenamefont {Sukachev}, \citenamefont {Bhaskar}, \citenamefont {Machielse},
  \citenamefont {Levonian}, \citenamefont {Knall}, \citenamefont {Stroganov},
  \citenamefont {Riedinger}, \citenamefont {Park}, \citenamefont {Lon{\v
  c}ar},\ and\ \citenamefont {Lukin}}]{Nguyen2019}%
  \BibitemOpen
  \bibfield  {author} {\bibinfo {author} {\bibfnamefont {C.~T.}\ \bibnamefont
  {Nguyen}}, \bibinfo {author} {\bibfnamefont {D.~D.}\ \bibnamefont
  {Sukachev}}, \bibinfo {author} {\bibfnamefont {M.~K.}\ \bibnamefont
  {Bhaskar}}, \bibinfo {author} {\bibfnamefont {B.}~\bibnamefont {Machielse}},
  \bibinfo {author} {\bibfnamefont {D.~S.}\ \bibnamefont {Levonian}}, \bibinfo
  {author} {\bibfnamefont {E.~N.}\ \bibnamefont {Knall}}, \bibinfo {author}
  {\bibfnamefont {P.}~\bibnamefont {Stroganov}}, \bibinfo {author}
  {\bibfnamefont {R.}~\bibnamefont {Riedinger}}, \bibinfo {author}
  {\bibfnamefont {H.}~\bibnamefont {Park}}, \bibinfo {author} {\bibfnamefont
  {M.}~\bibnamefont {Lon{\v c}ar}}, \ and\ \bibinfo {author} {\bibfnamefont
  {M.~D.}\ \bibnamefont {Lukin}},\ }\bibfield  {title} {\enquote {\bibinfo
  {title} {{Quantum {{Network Nodes Based}} on {{Diamond Qubits}} with an
  {{Efficient Nanophotonic Interface}}}},}\ }\href {\doibase
  10.1103/PhysRevLett.123.183602} {\bibfield  {journal} {\bibinfo  {journal}
  {Phys. Rev. Lett.}\ }\textbf {\bibinfo {volume} {123}},\ \bibinfo {pages}
  {183602} (\bibinfo {year} {2019})}\BibitemShut {NoStop}%
\bibitem [{\citenamefont {{Gonz{\'a}lez-Tudela}}\ \emph
  {et~al.}(2015)\citenamefont {{Gonz{\'a}lez-Tudela}}, \citenamefont {Hung},
  \citenamefont {Chang}, \citenamefont {Cirac},\ and\ \citenamefont
  {Kimble}}]{Gonzalez-Tudela2015}%
  \BibitemOpen
  \bibfield  {author} {\bibinfo {author} {\bibfnamefont {A.}~\bibnamefont
  {{Gonz{\'a}lez-Tudela}}}, \bibinfo {author} {\bibfnamefont {C.-L.}\
  \bibnamefont {Hung}}, \bibinfo {author} {\bibfnamefont {D.~E.}\ \bibnamefont
  {Chang}}, \bibinfo {author} {\bibfnamefont {J.~I.}\ \bibnamefont {Cirac}}, \
  and\ \bibinfo {author} {\bibfnamefont {H.~J.}\ \bibnamefont {Kimble}},\
  }\bibfield  {title} {\enquote {\bibinfo {title} {{Subwavelength Vacuum
  Lattices and Atom\textendash Atom Interactions in Two-Dimensional Photonic
  Crystals}},}\ }\href {\doibase 10.1038/nphoton.2015.54} {\bibfield  {journal}
  {\bibinfo  {journal} {Nat. Photonics}\ }\textbf {\bibinfo {volume} {9}},\
  \bibinfo {pages} {320} (\bibinfo {year} {2015})}\BibitemShut {NoStop}%
\bibitem [{\citenamefont {Douglas}\ \emph {et~al.}(2015)\citenamefont
  {Douglas}, \citenamefont {Habibian}, \citenamefont {Hung}, \citenamefont
  {Gorshkov}, \citenamefont {Kimble},\ and\ \citenamefont
  {Chang}}]{Douglas2015}%
  \BibitemOpen
  \bibfield  {author} {\bibinfo {author} {\bibfnamefont {J.~S.}\ \bibnamefont
  {Douglas}}, \bibinfo {author} {\bibfnamefont {H.}~\bibnamefont {Habibian}},
  \bibinfo {author} {\bibfnamefont {C.-L.}\ \bibnamefont {Hung}}, \bibinfo
  {author} {\bibfnamefont {A.~V.}\ \bibnamefont {Gorshkov}}, \bibinfo {author}
  {\bibfnamefont {H.~J.}\ \bibnamefont {Kimble}}, \ and\ \bibinfo {author}
  {\bibfnamefont {D.~E.}\ \bibnamefont {Chang}},\ }\bibfield  {title} {\enquote
  {\bibinfo {title} {{Quantum Many-Body Models with Cold Atoms Coupled to
  Photonic Crystals}},}\ }\href {\doibase 10.1038/nphoton.2015.57} {\bibfield
  {journal} {\bibinfo  {journal} {Nat. Photonics}\ }\textbf {\bibinfo {volume}
  {9}},\ \bibinfo {pages} {326} (\bibinfo {year} {2015})}\BibitemShut {NoStop}%
\bibitem [{\citenamefont {Khurgin}(2015)}]{Khurgin2015}%
  \BibitemOpen
  \bibfield  {author} {\bibinfo {author} {\bibfnamefont {J.~B.}\ \bibnamefont
  {Khurgin}},\ }\bibfield  {title} {\enquote {\bibinfo {title} {{How to Deal
  with the Loss in Plasmonics and Metamaterials}},}\ }\href {\doibase
  10.1038/nnano.2014.310} {\bibfield  {journal} {\bibinfo  {journal} {Nat.
  Nanotechnol.}\ }\textbf {\bibinfo {volume} {10}},\ \bibinfo {pages} {2}
  (\bibinfo {year} {2015})}\BibitemShut {NoStop}%
\bibitem [{\citenamefont {Li}\ \emph {et~al.}(2016)\citenamefont {Li},
  \citenamefont {{Hern{\'a}ngomez-P{\'e}rez}}, \citenamefont
  {{Garc{\'i}a-Vidal}},\ and\ \citenamefont
  {{Fern{\'a}ndez-Dom{\'i}nguez}}}]{Li2016Transformation}%
  \BibitemOpen
  \bibfield  {author} {\bibinfo {author} {\bibfnamefont {R.-Q.}\ \bibnamefont
  {Li}}, \bibinfo {author} {\bibfnamefont {D.}~\bibnamefont
  {{Hern{\'a}ngomez-P{\'e}rez}}}, \bibinfo {author} {\bibfnamefont {F.~J.}\
  \bibnamefont {{Garc{\'i}a-Vidal}}}, \ and\ \bibinfo {author} {\bibfnamefont
  {A.~I.}\ \bibnamefont {{Fern{\'a}ndez-Dom{\'i}nguez}}},\ }\bibfield  {title}
  {\enquote {\bibinfo {title} {{Transformation {{Optics Approach}} to
  {{Plasmon-Exciton Strong Coupling}} in {{Nanocavities}}}},}\ }\href {\doibase
  10.1103/PhysRevLett.117.107401} {\bibfield  {journal} {\bibinfo  {journal}
  {Phys. Rev. Lett.}\ }\textbf {\bibinfo {volume} {117}},\ \bibinfo {pages}
  {107401} (\bibinfo {year} {2016})}\BibitemShut {NoStop}%
\bibitem [{\citenamefont {Glauber}\ and\ \citenamefont
  {Lewenstein}(1991)}]{Glauber1991}%
  \BibitemOpen
  \bibfield  {author} {\bibinfo {author} {\bibfnamefont {R.~J.}\ \bibnamefont
  {Glauber}}\ and\ \bibinfo {author} {\bibfnamefont {M.}~\bibnamefont
  {Lewenstein}},\ }\bibfield  {title} {\enquote {\bibinfo {title} {{Quantum
  Optics of Dielectric Media}},}\ }\href {\doibase 10.1103/PhysRevA.43.467}
  {\bibfield  {journal} {\bibinfo  {journal} {Phys. Rev. A}\ }\textbf {\bibinfo
  {volume} {43}},\ \bibinfo {pages} {467} (\bibinfo {year} {1991})}\BibitemShut
  {NoStop}%
\bibitem [{\citenamefont {Cohen-Tannoudji}\ \emph {et~al.}(1997)\citenamefont
  {Cohen-Tannoudji}, \citenamefont {Dupont-Roc},\ and\ \citenamefont
  {Grynberg}}]{Cohen-Tannoudji1997}%
  \BibitemOpen
  \bibfield  {author} {\bibinfo {author} {\bibfnamefont {C.}~\bibnamefont
  {Cohen-Tannoudji}}, \bibinfo {author} {\bibfnamefont {J.}~\bibnamefont
  {Dupont-Roc}}, \ and\ \bibinfo {author} {\bibfnamefont {G.}~\bibnamefont
  {Grynberg}},\ }\href {\doibase 10.1002/9783527618422} {\emph {\bibinfo
  {title} {{Photons and {{Atoms}}: {{Introduction}} to {{Quantum
  Electrodynamics}}}}}},\ \bibinfo {edition} {1st}\ ed.\ (\bibinfo  {publisher}
  {{Wiley}},\ \bibinfo {year} {1997})\BibitemShut {NoStop}%
\bibitem [{\citenamefont {Fano}(1956)}]{Fano1956}%
  \BibitemOpen
  \bibfield  {author} {\bibinfo {author} {\bibfnamefont {U.}~\bibnamefont
  {Fano}},\ }\bibfield  {title} {\enquote {\bibinfo {title} {{Atomic {{Theory}}
  of {{Electromagnetic Interactions}} in {{Dense Materials}}}},}\ }\href
  {\doibase 10.1103/PhysRev.103.1202} {\bibfield  {journal} {\bibinfo
  {journal} {Phys. Rev.}\ }\textbf {\bibinfo {volume} {103}},\ \bibinfo {pages}
  {1202} (\bibinfo {year} {1956})}\BibitemShut {NoStop}%
\bibitem [{\citenamefont {Huttner}\ and\ \citenamefont
  {Barnett}(1992)}]{Huttner1992}%
  \BibitemOpen
  \bibfield  {author} {\bibinfo {author} {\bibfnamefont {B.}~\bibnamefont
  {Huttner}}\ and\ \bibinfo {author} {\bibfnamefont {S.~M.}\ \bibnamefont
  {Barnett}},\ }\bibfield  {title} {\enquote {\bibinfo {title} {{Quantization
  of the Electromagnetic Field in Dielectrics}},}\ }\href {\doibase
  10.1103/PhysRevA.46.4306} {\bibfield  {journal} {\bibinfo  {journal} {Phys.
  Rev. A}\ }\textbf {\bibinfo {volume} {46}},\ \bibinfo {pages} {4306}
  (\bibinfo {year} {1992})}\BibitemShut {NoStop}%
\bibitem [{\citenamefont {Scheel}\ \emph {et~al.}(1998)\citenamefont {Scheel},
  \citenamefont {Kn{\"o}ll},\ and\ \citenamefont {Welsch}}]{Scheel1998}%
  \BibitemOpen
  \bibfield  {author} {\bibinfo {author} {\bibfnamefont {S.}~\bibnamefont
  {Scheel}}, \bibinfo {author} {\bibfnamefont {L.}~\bibnamefont {Kn{\"o}ll}}, \
  and\ \bibinfo {author} {\bibfnamefont {D.-G.}\ \bibnamefont {Welsch}},\
  }\bibfield  {title} {\enquote {\bibinfo {title} {{{{QED}} Commutation
  Relations for Inhomogeneous {{Kramers-Kronig}} Dielectrics}},}\ }\href
  {\doibase 10.1103/PhysRevA.58.700} {\bibfield  {journal} {\bibinfo  {journal}
  {Phys. Rev. A}\ }\textbf {\bibinfo {volume} {58}},\ \bibinfo {pages} {700}
  (\bibinfo {year} {1998})}\BibitemShut {NoStop}%
\bibitem [{\citenamefont {Kn{\"o}ll}\ \emph {et~al.}(2001)\citenamefont
  {Kn{\"o}ll}, \citenamefont {Scheel},\ and\ \citenamefont
  {Welsch}}]{Knoll2001}%
  \BibitemOpen
  \bibfield  {author} {\bibinfo {author} {\bibfnamefont {L.}~\bibnamefont
  {Kn{\"o}ll}}, \bibinfo {author} {\bibfnamefont {S.}~\bibnamefont {Scheel}}, \
  and\ \bibinfo {author} {\bibfnamefont {D.-G.}\ \bibnamefont {Welsch}},\
  }\bibfield  {title} {\enquote {\bibinfo {title} {{{{QED}} in {{Dispersing}}
  and {{Absorbing Media}}}},}\ }in\ \href@noop {} {\emph {\bibinfo {booktitle}
  {Coherence and {{Statistics}} of {{Photons}} and {{Atoms}}}}},\ \bibinfo
  {editor} {edited by\ \bibinfo {editor} {\bibfnamefont {J.}~\bibnamefont
  {Pe{\v r}ina}}}\ (\bibinfo  {publisher} {{WILEY-VCH Verlag}},\ \bibinfo
  {address} {{New York}},\ \bibinfo {year} {2001})\ \bibinfo {edition} {1st}\
  ed.,\ \Eprint {http://arxiv.org/abs/quant-ph/0006121}
  {arXiv:quant-ph/0006121} \BibitemShut {NoStop}%
\bibitem [{\citenamefont {Scheel}\ and\ \citenamefont
  {Buhmann}(2008)}]{Scheel2008}%
  \BibitemOpen
  \bibfield  {author} {\bibinfo {author} {\bibfnamefont {S.}~\bibnamefont
  {Scheel}}\ and\ \bibinfo {author} {\bibfnamefont {S.~Y.}\ \bibnamefont
  {Buhmann}},\ }\bibfield  {title} {\enquote {\bibinfo {title} {{Macroscopic
  Quantum Electrodynamics - {{Concepts}} and Applications}},}\ }\href {\doibase
  10.2478/v10155-010-0092-x} {\bibfield  {journal} {\bibinfo  {journal} {Acta
  Phys. Slovaca}\ }\textbf {\bibinfo {volume} {58}},\ \bibinfo {pages} {675}
  (\bibinfo {year} {2008})}\BibitemShut {NoStop}%
\bibitem [{\citenamefont {Buhmann}(2012{\natexlab{a}})}]{Buhmann2012I}%
  \BibitemOpen
  \bibfield  {author} {\bibinfo {author} {\bibfnamefont {S.~Y.}\ \bibnamefont
  {Buhmann}},\ }\href {\doibase 10.1007/978-3-642-32484-0} {\emph {\bibinfo
  {title} {{Dispersion {{Forces I}}}}}},\ \bibinfo {series} {Springer
  {{Tracts}} in {{Modern Physics}}}, Vol.\ \bibinfo {volume} {247}\ (\bibinfo
  {publisher} {{Springer Berlin Heidelberg}},\ \bibinfo {address} {{Berlin,
  Heidelberg}},\ \bibinfo {year} {2012})\BibitemShut {NoStop}%
\bibitem [{\citenamefont {Buhmann}(2012{\natexlab{b}})}]{Buhmann2012II}%
  \BibitemOpen
  \bibfield  {author} {\bibinfo {author} {\bibfnamefont {S.~Y.}\ \bibnamefont
  {Buhmann}},\ }\href {\doibase 10.1007/978-3-642-32466-6} {\emph {\bibinfo
  {title} {{Dispersion {{Forces II}}}}}},\ \bibinfo {series} {Springer
  {{Tracts}} in {{Modern Physics}}}, Vol.\ \bibinfo {volume} {248}\ (\bibinfo
  {publisher} {{Springer Berlin Heidelberg}},\ \bibinfo {address} {{Berlin,
  Heidelberg}},\ \bibinfo {year} {2012})\BibitemShut {NoStop}%
\bibitem [{\citenamefont {Feist}\ \emph {et~al.}(2020)\citenamefont {Feist},
  \citenamefont {{Fern{\'a}ndez-Dom{\'i}nguez}},\ and\ \citenamefont
  {{Garc{\'i}a-Vidal}}}]{Feist2020}%
  \BibitemOpen
  \bibfield  {author} {\bibinfo {author} {\bibfnamefont {J.}~\bibnamefont
  {Feist}}, \bibinfo {author} {\bibfnamefont {A.~I.}\ \bibnamefont
  {{Fern{\'a}ndez-Dom{\'i}nguez}}}, \ and\ \bibinfo {author} {\bibfnamefont
  {F.~J.}\ \bibnamefont {{Garc{\'i}a-Vidal}}},\ }\bibfield  {title} {\enquote
  {\bibinfo {title} {{Macroscopic {{QED}} for Quantum Nanophotonics:
  Emitter-Centered Modes as a Minimal Basis for Multiemitter Problems}},}\
  }\href {\doibase 10.1515/nanoph-2020-0451} {\bibfield  {journal} {\bibinfo
  {journal} {Nanophotonics}\ }\textbf {\bibinfo {volume} {10}},\ \bibinfo
  {pages} {477} (\bibinfo {year} {2020})}\BibitemShut {NoStop}%
\bibitem [{\citenamefont {Sauvan}\ \emph {et~al.}(2013)\citenamefont {Sauvan},
  \citenamefont {Hugonin}, \citenamefont {Maksymov},\ and\ \citenamefont
  {Lalanne}}]{Sauvan2013}%
  \BibitemOpen
  \bibfield  {author} {\bibinfo {author} {\bibfnamefont {C.}~\bibnamefont
  {Sauvan}}, \bibinfo {author} {\bibfnamefont {J.~P.}\ \bibnamefont {Hugonin}},
  \bibinfo {author} {\bibfnamefont {I.~S.}\ \bibnamefont {Maksymov}}, \ and\
  \bibinfo {author} {\bibfnamefont {P.}~\bibnamefont {Lalanne}},\ }\bibfield
  {title} {\enquote {\bibinfo {title} {{Theory of the {{Spontaneous Optical
  Emission}} of {{Nanosize Photonic}} and {{Plasmon Resonators}}}},}\ }\href
  {\doibase 10.1103/PhysRevLett.110.237401} {\bibfield  {journal} {\bibinfo
  {journal} {Phys. Rev. Lett.}\ }\textbf {\bibinfo {volume} {110}},\ \bibinfo
  {pages} {237401} (\bibinfo {year} {2013})}\BibitemShut {NoStop}%
\bibitem [{\citenamefont {Kristensen}\ and\ \citenamefont
  {Hughes}(2014)}]{Kristensen2014}%
  \BibitemOpen
  \bibfield  {author} {\bibinfo {author} {\bibfnamefont {P.~T.}\ \bibnamefont
  {Kristensen}}\ and\ \bibinfo {author} {\bibfnamefont {S.}~\bibnamefont
  {Hughes}},\ }\bibfield  {title} {\enquote {\bibinfo {title} {{Modes and
  {{Mode Volumes}} of {{Leaky Optical Cavities}} and {{Plasmonic
  Nanoresonators}}}},}\ }\href {\doibase 10.1021/ph400114e} {\bibfield
  {journal} {\bibinfo  {journal} {ACS Photonics}\ }\textbf {\bibinfo {volume}
  {1}},\ \bibinfo {pages} {2} (\bibinfo {year} {2014})}\BibitemShut {NoStop}%
\bibitem [{\citenamefont {Lalanne}\ \emph {et~al.}(2018)\citenamefont
  {Lalanne}, \citenamefont {Yan}, \citenamefont {Vynck}, \citenamefont
  {Sauvan},\ and\ \citenamefont {Hugonin}}]{Lalanne2018}%
  \BibitemOpen
  \bibfield  {author} {\bibinfo {author} {\bibfnamefont {P.}~\bibnamefont
  {Lalanne}}, \bibinfo {author} {\bibfnamefont {W.}~\bibnamefont {Yan}},
  \bibinfo {author} {\bibfnamefont {K.}~\bibnamefont {Vynck}}, \bibinfo
  {author} {\bibfnamefont {C.}~\bibnamefont {Sauvan}}, \ and\ \bibinfo {author}
  {\bibfnamefont {J.-P.}\ \bibnamefont {Hugonin}},\ }\bibfield  {title}
  {\enquote {\bibinfo {title} {{Light {{Interaction}} with {{Photonic}} and
  {{Plasmonic Resonances}}}},}\ }\href {\doibase 10.1002/lpor.201700113}
  {\bibfield  {journal} {\bibinfo  {journal} {Laser Photonics Rev.}\ }\textbf
  {\bibinfo {volume} {12}},\ \bibinfo {pages} {1700113} (\bibinfo {year}
  {2018})}\BibitemShut {NoStop}%
\bibitem [{\citenamefont {Franke}\ \emph {et~al.}(2019)\citenamefont {Franke},
  \citenamefont {Hughes}, \citenamefont {Kamandar~Dezfouli}, \citenamefont
  {Kristensen}, \citenamefont {Busch}, \citenamefont {Knorr},\ and\
  \citenamefont {Richter}}]{Franke2019}%
  \BibitemOpen
  \bibfield  {author} {\bibinfo {author} {\bibfnamefont {S.}~\bibnamefont
  {Franke}}, \bibinfo {author} {\bibfnamefont {S.}~\bibnamefont {Hughes}},
  \bibinfo {author} {\bibfnamefont {M.}~\bibnamefont {Kamandar~Dezfouli}},
  \bibinfo {author} {\bibfnamefont {P.~T.}\ \bibnamefont {Kristensen}},
  \bibinfo {author} {\bibfnamefont {K.}~\bibnamefont {Busch}}, \bibinfo
  {author} {\bibfnamefont {A.}~\bibnamefont {Knorr}}, \ and\ \bibinfo {author}
  {\bibfnamefont {M.}~\bibnamefont {Richter}},\ }\bibfield  {title} {\enquote
  {\bibinfo {title} {{Quantization of {{Quasinormal Modes}} for {{Open
  Cavities}} and {{Plasmonic Cavity Quantum Electrodynamics}}}},}\ }\href
  {\doibase 10.1103/PhysRevLett.122.213901} {\bibfield  {journal} {\bibinfo
  {journal} {Phys. Rev. Lett.}\ }\textbf {\bibinfo {volume} {122}},\ \bibinfo
  {pages} {213901} (\bibinfo {year} {2019})}\BibitemShut {NoStop}%
\bibitem [{\citenamefont {Ren}\ \emph {et~al.}(2021)\citenamefont {Ren},
  \citenamefont {Franke},\ and\ \citenamefont {Hughes}}]{Ren2021}%
  \BibitemOpen
  \bibfield  {author} {\bibinfo {author} {\bibfnamefont {J.}~\bibnamefont
  {Ren}}, \bibinfo {author} {\bibfnamefont {S.}~\bibnamefont {Franke}}, \ and\
  \bibinfo {author} {\bibfnamefont {S.}~\bibnamefont {Hughes}},\ }\bibfield
  {title} {\enquote {\bibinfo {title} {{Quasinormal {{Modes}}, {{Local
  Density}} of {{States}}, and {{Classical Purcell Factors}} for {{Coupled
  Loss-Gain Resonators}}}},}\ }\href {\doibase 10.1103/PhysRevX.11.041020}
  {\bibfield  {journal} {\bibinfo  {journal} {Phys. Rev. X}\ }\textbf {\bibinfo
  {volume} {11}},\ \bibinfo {pages} {041020} (\bibinfo {year}
  {2021})}\BibitemShut {NoStop}%
\bibitem [{\citenamefont {Franke}\ \emph {et~al.}()\citenamefont {Franke},
  \citenamefont {Ren},\ and\ \citenamefont {Hughes}}]{Franke2021}%
  \BibitemOpen
  \bibfield  {author} {\bibinfo {author} {\bibfnamefont {S.}~\bibnamefont
  {Franke}}, \bibinfo {author} {\bibfnamefont {J.}~\bibnamefont {Ren}}, \ and\
  \bibinfo {author} {\bibfnamefont {S.}~\bibnamefont {Hughes}},\ }\href@noop {}
  {\enquote {\bibinfo {title} {{Quantized Quasinormal Mode Theory of Coupled
  Lossy and Amplifying Resonators}},}\ }\Eprint
  {http://arxiv.org/abs/2109.14967} {arXiv:2109.14967} \BibitemShut {NoStop}%
\bibitem [{\citenamefont {Delga}\ \emph {et~al.}(2014)\citenamefont {Delga},
  \citenamefont {Feist}, \citenamefont {{Bravo-Abad}},\ and\ \citenamefont
  {{Garcia-Vidal}}}]{Delga2014}%
  \BibitemOpen
  \bibfield  {author} {\bibinfo {author} {\bibfnamefont {A.}~\bibnamefont
  {Delga}}, \bibinfo {author} {\bibfnamefont {J.}~\bibnamefont {Feist}},
  \bibinfo {author} {\bibfnamefont {J.}~\bibnamefont {{Bravo-Abad}}}, \ and\
  \bibinfo {author} {\bibfnamefont {F.~J.}\ \bibnamefont {{Garcia-Vidal}}},\
  }\bibfield  {title} {\enquote {\bibinfo {title} {{Quantum {{Emitters Near}} a
  {{Metal Nanoparticle}}: {{Strong Coupling}} and {{Quenching}}}},}\ }\href
  {\doibase 10.1103/PhysRevLett.112.253601} {\bibfield  {journal} {\bibinfo
  {journal} {Phys. Rev. Lett.}\ }\textbf {\bibinfo {volume} {112}},\ \bibinfo
  {pages} {253601} (\bibinfo {year} {2014})}\BibitemShut {NoStop}%
\bibitem [{\citenamefont {{Cuartero-Gonz{\'a}lez}}\ and\ \citenamefont
  {{Fern{\'a}ndez-Dom{\'i}nguez}}(2020)}]{Cuartero-Gonzalez2020Dipolar}%
  \BibitemOpen
  \bibfield  {author} {\bibinfo {author} {\bibfnamefont {A.}~\bibnamefont
  {{Cuartero-Gonz{\'a}lez}}}\ and\ \bibinfo {author} {\bibfnamefont {A.~I.}\
  \bibnamefont {{Fern{\'a}ndez-Dom{\'i}nguez}}},\ }\bibfield  {title} {\enquote
  {\bibinfo {title} {{Dipolar and Quadrupolar Excitons Coupled to a
  Nanoparticle-on-Mirror Cavity}},}\ }\href {\doibase
  10.1103/PhysRevB.101.035403} {\bibfield  {journal} {\bibinfo  {journal}
  {Phys. Rev. B}\ }\textbf {\bibinfo {volume} {101}},\ \bibinfo {pages}
  {035403} (\bibinfo {year} {2020})}\BibitemShut {NoStop}%
\bibitem [{\citenamefont {Fregoni}\ \emph {et~al.}(2021)\citenamefont
  {Fregoni}, \citenamefont {Haugland}, \citenamefont {Pipolo}, \citenamefont
  {Giovannini}, \citenamefont {Koch},\ and\ \citenamefont
  {Corni}}]{Fregoni2021Strong}%
  \BibitemOpen
  \bibfield  {author} {\bibinfo {author} {\bibfnamefont {J.}~\bibnamefont
  {Fregoni}}, \bibinfo {author} {\bibfnamefont {T.~S.}\ \bibnamefont
  {Haugland}}, \bibinfo {author} {\bibfnamefont {S.}~\bibnamefont {Pipolo}},
  \bibinfo {author} {\bibfnamefont {T.}~\bibnamefont {Giovannini}}, \bibinfo
  {author} {\bibfnamefont {H.}~\bibnamefont {Koch}}, \ and\ \bibinfo {author}
  {\bibfnamefont {S.}~\bibnamefont {Corni}},\ }\bibfield  {title} {\enquote
  {\bibinfo {title} {{Strong {{Coupling}} between {{Localized Surface
  Plasmons}} and {{Molecules}} by {{Coupled Cluster Theory}}}},}\ }\href
  {\doibase 10.1021/acs.nanolett.1c02162} {\bibfield  {journal} {\bibinfo
  {journal} {Nano Lett.}\ }\textbf {\bibinfo {volume} {21}},\ \bibinfo {pages}
  {6664} (\bibinfo {year} {2021})}\BibitemShut {NoStop}%
\bibitem [{\citenamefont {{Gonz{\'a}lez-Tudela}}\ \emph
  {et~al.}(2014)\citenamefont {{Gonz{\'a}lez-Tudela}}, \citenamefont
  {Huidobro}, \citenamefont {{Mart{\'i}n-Moreno}}, \citenamefont {Tejedor},\
  and\ \citenamefont {{Garc{\'i}a-Vidal}}}]{Gonzalez-Tudela2014}%
  \BibitemOpen
  \bibfield  {author} {\bibinfo {author} {\bibfnamefont {A.}~\bibnamefont
  {{Gonz{\'a}lez-Tudela}}}, \bibinfo {author} {\bibfnamefont {P.~A.}\
  \bibnamefont {Huidobro}}, \bibinfo {author} {\bibfnamefont {L.}~\bibnamefont
  {{Mart{\'i}n-Moreno}}}, \bibinfo {author} {\bibfnamefont {C.}~\bibnamefont
  {Tejedor}}, \ and\ \bibinfo {author} {\bibfnamefont {F.~J.}\ \bibnamefont
  {{Garc{\'i}a-Vidal}}},\ }\bibfield  {title} {\enquote {\bibinfo {title}
  {{Reversible Dynamics of Single Quantum Emitters near Metal-Dielectric
  Interfaces}},}\ }\href {\doibase 10.1103/PhysRevB.89.041402} {\bibfield
  {journal} {\bibinfo  {journal} {Phys. Rev. B}\ }\textbf {\bibinfo {volume}
  {89}},\ \bibinfo {pages} {041402(R)} (\bibinfo {year} {2014})}\BibitemShut
  {NoStop}%
\bibitem [{\citenamefont {{Cuartero-Gonz{\'a}lez}}\ and\ \citenamefont
  {{Fern{\'a}ndez-Dom{\'i}nguez}}(2018)}]{Cuartero-Gonzalez2018}%
  \BibitemOpen
  \bibfield  {author} {\bibinfo {author} {\bibfnamefont {A.}~\bibnamefont
  {{Cuartero-Gonz{\'a}lez}}}\ and\ \bibinfo {author} {\bibfnamefont {A.~I.}\
  \bibnamefont {{Fern{\'a}ndez-Dom{\'i}nguez}}},\ }\bibfield  {title} {\enquote
  {\bibinfo {title} {{Light-{{Forbidden Transitions}} in {{Plasmon-Emitter
  Interactions}} beyond the {{Weak Coupling Regime}}}},}\ }\href {\doibase
  10.1021/acsphotonics.8b00678} {\bibfield  {journal} {\bibinfo  {journal} {ACS
  Photonics}\ }\textbf {\bibinfo {volume} {5}},\ \bibinfo {pages} {3415}
  (\bibinfo {year} {2018})}\BibitemShut {NoStop}%
\bibitem [{\citenamefont {Yang}\ \emph {et~al.}(2019)\citenamefont {Yang},
  \citenamefont {An},\ and\ \citenamefont {Lin}}]{Yang2019Signatures}%
  \BibitemOpen
  \bibfield  {author} {\bibinfo {author} {\bibfnamefont {C.-J.}\ \bibnamefont
  {Yang}}, \bibinfo {author} {\bibfnamefont {J.-H.}\ \bibnamefont {An}}, \ and\
  \bibinfo {author} {\bibfnamefont {H.-Q.}\ \bibnamefont {Lin}},\ }\bibfield
  {title} {\enquote {\bibinfo {title} {{Signatures of Quantized Coupling
  between Quantum Emitters and Localized Surface Plasmons}},}\ }\href {\doibase
  10.1103/PhysRevResearch.1.023027} {\bibfield  {journal} {\bibinfo  {journal}
  {Phys. Rev. Research}\ }\textbf {\bibinfo {volume} {1}},\ \bibinfo {pages}
  {023027} (\bibinfo {year} {2019})}\BibitemShut {NoStop}%
\bibitem [{\citenamefont {{Cuartero-Gonz{\'a}lez}}\ \emph
  {et~al.}(2021)\citenamefont {{Cuartero-Gonz{\'a}lez}}, \citenamefont
  {Manjavacas},\ and\ \citenamefont
  {{Fern{\'a}ndez-Dom{\'i}nguez}}}]{Cuartero-Gonzalez2021Distortion}%
  \BibitemOpen
  \bibfield  {author} {\bibinfo {author} {\bibfnamefont {A.}~\bibnamefont
  {{Cuartero-Gonz{\'a}lez}}}, \bibinfo {author} {\bibfnamefont
  {A.}~\bibnamefont {Manjavacas}}, \ and\ \bibinfo {author} {\bibfnamefont
  {A.~I.}\ \bibnamefont {{Fern{\'a}ndez-Dom{\'i}nguez}}},\ }\bibfield  {title}
  {\enquote {\bibinfo {title} {{Distortion of the Local Density of States in a
  Plasmonic Cavity by a Quantum Emitter}},}\ }\href {\doibase
  10.1088/1367-2630/ac0199} {\bibfield  {journal} {\bibinfo  {journal} {New J.
  Phys.}\ }\textbf {\bibinfo {volume} {23}},\ \bibinfo {pages} {073011}
  (\bibinfo {year} {2021})}\BibitemShut {NoStop}%
\bibitem [{\citenamefont {Breuer}\ and\ \citenamefont
  {Petruccione}(2007)}]{Breuer2007}%
  \BibitemOpen
  \bibfield  {author} {\bibinfo {author} {\bibfnamefont {H.-P.}\ \bibnamefont
  {Breuer}}\ and\ \bibinfo {author} {\bibfnamefont {F.}~\bibnamefont
  {Petruccione}},\ }\href {\doibase 10.1093/acprof:oso/9780199213900.001.0001}
  {\emph {\bibinfo {title} {{The {{Theory}} of {{Open Quantum Systems}}}}}}\
  (\bibinfo  {publisher} {{Oxford University Press}},\ \bibinfo {year}
  {2007})\BibitemShut {NoStop}%
\bibitem [{\citenamefont {Mascherpa}\ \emph {et~al.}(2020)\citenamefont
  {Mascherpa}, \citenamefont {Smirne}, \citenamefont {Somoza}, \citenamefont
  {{Fern{\'a}ndez-Acebal}}, \citenamefont {Donadi}, \citenamefont {Tamascelli},
  \citenamefont {Huelga},\ and\ \citenamefont {Plenio}}]{Mascherpa2020}%
  \BibitemOpen
  \bibfield  {author} {\bibinfo {author} {\bibfnamefont {F.}~\bibnamefont
  {Mascherpa}}, \bibinfo {author} {\bibfnamefont {A.}~\bibnamefont {Smirne}},
  \bibinfo {author} {\bibfnamefont {A.~D.}\ \bibnamefont {Somoza}}, \bibinfo
  {author} {\bibfnamefont {P.}~\bibnamefont {{Fern{\'a}ndez-Acebal}}}, \bibinfo
  {author} {\bibfnamefont {S.}~\bibnamefont {Donadi}}, \bibinfo {author}
  {\bibfnamefont {D.}~\bibnamefont {Tamascelli}}, \bibinfo {author}
  {\bibfnamefont {S.~F.}\ \bibnamefont {Huelga}}, \ and\ \bibinfo {author}
  {\bibfnamefont {M.~B.}\ \bibnamefont {Plenio}},\ }\bibfield  {title}
  {\enquote {\bibinfo {title} {{Optimized Auxiliary Oscillators for the
  Simulation of General Open Quantum Systems}},}\ }\href {\doibase
  10.1103/PhysRevA.101.052108} {\bibfield  {journal} {\bibinfo  {journal}
  {Phys. Rev. A}\ }\textbf {\bibinfo {volume} {101}},\ \bibinfo {pages}
  {052108} (\bibinfo {year} {2020})}\BibitemShut {NoStop}%
\bibitem [{\citenamefont {Glutsch}(2002)}]{Glutsch2002}%
  \BibitemOpen
  \bibfield  {author} {\bibinfo {author} {\bibfnamefont {S.}~\bibnamefont
  {Glutsch}},\ }\bibfield  {title} {\enquote {\bibinfo {title} {{Optical
  Absorption of the {{Fano}} Model: {{General}} Case of Many Resonances and
  Many Continua}},}\ }\href {\doibase 10.1103/PhysRevB.66.075310} {\bibfield
  {journal} {\bibinfo  {journal} {Phys. Rev. B}\ }\textbf {\bibinfo {volume}
  {66}},\ \bibinfo {pages} {075310} (\bibinfo {year} {2002})}\BibitemShut
  {NoStop}%
\bibitem [{\citenamefont {Lentrodt}\ \emph {et~al.}()\citenamefont {Lentrodt},
  \citenamefont {Diekmann}, \citenamefont {Keitel}, \citenamefont {Rotter},\
  and\ \citenamefont {Evers}}]{Lentrodt2021}%
  \BibitemOpen
  \bibfield  {author} {\bibinfo {author} {\bibfnamefont {D.}~\bibnamefont
  {Lentrodt}}, \bibinfo {author} {\bibfnamefont {O.}~\bibnamefont {Diekmann}},
  \bibinfo {author} {\bibfnamefont {C.~H.}\ \bibnamefont {Keitel}}, \bibinfo
  {author} {\bibfnamefont {S.}~\bibnamefont {Rotter}}, \ and\ \bibinfo {author}
  {\bibfnamefont {J.}~\bibnamefont {Evers}},\ }\href@noop {} {\enquote
  {\bibinfo {title} {{Classifying and Harnessing Multi-Mode Light-Matter
  Interaction in Lossy Resonators}},}\ }\Eprint
  {http://arxiv.org/abs/2107.11775} {arXiv:2107.11775} \BibitemShut {NoStop}%
\bibitem [{\citenamefont {{S{\'a}nchez-Barquilla}}\ and\ \citenamefont
  {Feist}(2021)}]{Sanchez-Barquilla2021}%
  \BibitemOpen
  \bibfield  {author} {\bibinfo {author} {\bibfnamefont {M.}~\bibnamefont
  {{S{\'a}nchez-Barquilla}}}\ and\ \bibinfo {author} {\bibfnamefont
  {J.}~\bibnamefont {Feist}},\ }\bibfield  {title} {\enquote {\bibinfo {title}
  {{Accurate {{Truncations}} of {{Chain Mapping Models}} for {{Open Quantum
  Systems}}}},}\ }\href {\doibase 10.3390/nano11082104} {\bibfield  {journal}
  {\bibinfo  {journal} {Nanomaterials}\ }\textbf {\bibinfo {volume} {11}},\
  \bibinfo {pages} {2104} (\bibinfo {year} {2021})}\BibitemShut {NoStop}%
\bibitem [{\citenamefont {Pleasance}\ \emph {et~al.}(2020)\citenamefont
  {Pleasance}, \citenamefont {Garraway},\ and\ \citenamefont
  {Petruccione}}]{Pleasance2020}%
  \BibitemOpen
  \bibfield  {author} {\bibinfo {author} {\bibfnamefont {G.}~\bibnamefont
  {Pleasance}}, \bibinfo {author} {\bibfnamefont {B.~M.}\ \bibnamefont
  {Garraway}}, \ and\ \bibinfo {author} {\bibfnamefont {F.}~\bibnamefont
  {Petruccione}},\ }\bibfield  {title} {\enquote {\bibinfo {title}
  {{Generalized Theory of Pseudomodes for Exact Descriptions of
  Non-{{Markovian}} Quantum Processes}},}\ }\href {\doibase
  10.1103/PhysRevResearch.2.043058} {\bibfield  {journal} {\bibinfo  {journal}
  {Phys. Rev. Research}\ }\textbf {\bibinfo {volume} {2}},\ \bibinfo {pages}
  {043058} (\bibinfo {year} {2020})}\BibitemShut {NoStop}%
\bibitem [{\citenamefont {Dung}\ \emph {et~al.}(2002)\citenamefont {Dung},
  \citenamefont {Kn{\"o}ll},\ and\ \citenamefont {Welsch}}]{Dung2002Resonant}%
  \BibitemOpen
  \bibfield  {author} {\bibinfo {author} {\bibfnamefont {H.~T.}\ \bibnamefont
  {Dung}}, \bibinfo {author} {\bibfnamefont {L.}~\bibnamefont {Kn{\"o}ll}}, \
  and\ \bibinfo {author} {\bibfnamefont {D.-G.}\ \bibnamefont {Welsch}},\
  }\bibfield  {title} {\enquote {\bibinfo {title} {{Resonant Dipole-Dipole
  Interaction in the Presence of Dispersing and Absorbing Surroundings}},}\
  }\href {\doibase 10.1103/PhysRevA.66.063810} {\bibfield  {journal} {\bibinfo
  {journal} {Phys. Rev. A}\ }\textbf {\bibinfo {volume} {66}},\ \bibinfo
  {pages} {063810} (\bibinfo {year} {2002})}\BibitemShut {NoStop}%
\bibitem [{\citenamefont {{Gonz{\'a}lez-Tudela}}\ \emph
  {et~al.}(2011)\citenamefont {{Gonz{\'a}lez-Tudela}}, \citenamefont
  {{Mart{\'i}n-Cano}}, \citenamefont {Moreno}, \citenamefont
  {{Mart{\'i}n-Moreno}}, \citenamefont {Tejedor},\ and\ \citenamefont
  {{Garc{\'i}a-Vidal}}}]{Gonzalez-Tudela2011}%
  \BibitemOpen
  \bibfield  {author} {\bibinfo {author} {\bibfnamefont {A.}~\bibnamefont
  {{Gonz{\'a}lez-Tudela}}}, \bibinfo {author} {\bibfnamefont {D.}~\bibnamefont
  {{Mart{\'i}n-Cano}}}, \bibinfo {author} {\bibfnamefont {E.}~\bibnamefont
  {Moreno}}, \bibinfo {author} {\bibfnamefont {L.}~\bibnamefont
  {{Mart{\'i}n-Moreno}}}, \bibinfo {author} {\bibfnamefont {C.}~\bibnamefont
  {Tejedor}}, \ and\ \bibinfo {author} {\bibfnamefont {F.~J.}\ \bibnamefont
  {{Garc{\'i}a-Vidal}}},\ }\bibfield  {title} {\enquote {\bibinfo {title}
  {{Entanglement of {{Two Qubits Mediated}} by {{One-Dimensional Plasmonic
  Waveguides}}}},}\ }\href {\doibase 10.1103/PhysRevLett.106.020501} {\bibfield
   {journal} {\bibinfo  {journal} {Phys. Rev. Lett.}\ }\textbf {\bibinfo
  {volume} {106}},\ \bibinfo {pages} {020501} (\bibinfo {year}
  {2011})}\BibitemShut {NoStop}%
\bibitem [{\citenamefont {{Mart{\'i}n-Cano}}\ \emph {et~al.}(2011)\citenamefont
  {{Mart{\'i}n-Cano}}, \citenamefont {{Gonz{\'a}lez-Tudela}}, \citenamefont
  {{Mart{\'i}n-Moreno}}, \citenamefont {{Garc{\'i}a-Vidal}}, \citenamefont
  {Tejedor},\ and\ \citenamefont {Moreno}}]{Martin-Cano2011}%
  \BibitemOpen
  \bibfield  {author} {\bibinfo {author} {\bibfnamefont {D.}~\bibnamefont
  {{Mart{\'i}n-Cano}}}, \bibinfo {author} {\bibfnamefont {A.}~\bibnamefont
  {{Gonz{\'a}lez-Tudela}}}, \bibinfo {author} {\bibfnamefont {L.}~\bibnamefont
  {{Mart{\'i}n-Moreno}}}, \bibinfo {author} {\bibfnamefont {F.~J.}\
  \bibnamefont {{Garc{\'i}a-Vidal}}}, \bibinfo {author} {\bibfnamefont
  {C.}~\bibnamefont {Tejedor}}, \ and\ \bibinfo {author} {\bibfnamefont
  {E.}~\bibnamefont {Moreno}},\ }\bibfield  {title} {\enquote {\bibinfo {title}
  {{Dissipation-Driven Generation of Two-Qubit Entanglement Mediated by
  Plasmonic Waveguides}},}\ }\href {\doibase 10.1103/PhysRevB.84.235306}
  {\bibfield  {journal} {\bibinfo  {journal} {Phys. Rev. B}\ }\textbf {\bibinfo
  {volume} {84}},\ \bibinfo {pages} {235306} (\bibinfo {year}
  {2011})}\BibitemShut {NoStop}%
\bibitem [{\citenamefont {Downing}\ \emph {et~al.}(2019)\citenamefont
  {Downing}, \citenamefont {L{\'o}pez~Carre{\~n}o}, \citenamefont {Laussy},
  \citenamefont {{del Valle}},\ and\ \citenamefont
  {{Fern{\'a}ndez-Dom{\'i}nguez}}}]{Downing2019}%
  \BibitemOpen
  \bibfield  {author} {\bibinfo {author} {\bibfnamefont {C.~A.}\ \bibnamefont
  {Downing}}, \bibinfo {author} {\bibfnamefont {J.~C.}\ \bibnamefont
  {L{\'o}pez~Carre{\~n}o}}, \bibinfo {author} {\bibfnamefont {F.~P.}\
  \bibnamefont {Laussy}}, \bibinfo {author} {\bibfnamefont {E.}~\bibnamefont
  {{del Valle}}}, \ and\ \bibinfo {author} {\bibfnamefont {A.~I.}\ \bibnamefont
  {{Fern{\'a}ndez-Dom{\'i}nguez}}},\ }\bibfield  {title} {\enquote {\bibinfo
  {title} {{Quasichiral {{Interactions}} between {{Quantum Emitters}} at the
  {{Nanoscale}}}},}\ }\href {\doibase 10.1103/PhysRevLett.122.057401}
  {\bibfield  {journal} {\bibinfo  {journal} {Phys. Rev. Lett.}\ }\textbf
  {\bibinfo {volume} {122}},\ \bibinfo {pages} {057401} (\bibinfo {year}
  {2019})}\BibitemShut {NoStop}%
\end{thebibliography}%

\end{document}